\documentclass[aps,prl,twocolumn,showpacs,superscriptaddress,groupedaddress]{revtex4}  % for review and submission
\usepackage{graphicx}  % needed for figures
\usepackage{dcolumn}   % needed for some tables
\usepackage{bm}        % for math
\usepackage{amssymb}   % for math

\begin{document}

% The following information is for internal review, please remove them for submission
\widetext
\leftline{Version xx as of \today}
\leftline{Primary authors: Michele Guala}
\leftline{To be submitted to (PRL)}
\leftline{Comment to {\tt mguala@umn.edu} by xxx, yyy}
\centerline{\em D\O\ INTERNAL DOCUMENT -- NOT FOR PUBLIC DISTRIBUTION}

% the following line is for submission, including submission to the arXiv!!
%\hspace{5.2in} \mbox{Fermilab-Pub-04/xxx-E}

\title{Particle growth in turbulent flow under dynamically critical Stokes conditions}
\author{Michele Guala}
\affiliation{St. Anthony Falls Laboratory,  University of Minnesota, Minneapolis, MN 55414, USA}
\affiliation{Civil, Environmental and Geo- Engineering, University of Minnesota, Minneapolis, MN 55455, USA\\}
\author{Jiarong Hong}
\affiliation{St. Anthony Falls Laboratory,  University of Minnesota, Minneapolis, MN 55414, USA}
\affiliation{Mechanical Engineering, University of Minnesota, Minneapolis, MN 55455, USA\\}
%
%\author{Michele Guala, Jiarong Hong}
%\affiliation{St. Anthony Falls Laboratory,  University of Minnesota, Minneapolis, MN 55414, USA\\}
%\affiliation{Civil, Environmental and Geo- Engineering, University of Minnesota, Minneapolis, MN 55414, USA\\}
%\affiliation{Mechanical Engineering, University of Minnesota, Minneapolis, MN 55414, USA}
\date{\today}
\begin{abstract}
A simple theory, based on observations of snowflake distribution in a turbulent flow, is proposed to model the growth of inertial particles as a result of dynamic clustering at scales larger than the Kolmogorov length scale. Particles able to stick or coalesce are expected to grow in size in flow regions where preferential concentration is predicted by a critical Stokes number $St=\tau_p/\tau_f \simeq 1 $. We postulate that, during growth, $St$ remains critical, with the particle response time $\tau_p$ evolving  according to the specific flow time scale $\tau_f$ defined by the vortices around which progressively larger particles end up orbiting. This mechanism leads to the prediction of the limiting size of aggregating particles in a turbulent flow. Such limit is determined by the extent of the turbulent inertial range, which can be formulated as a function of accessible integral-scale quantities. The proposed dynamically critical Stokes growth provides a framework to interpret particle aggregation, size growth and particle cluster growth in various geophysical multi-phase flows.
\end{abstract}
\maketitle
This study is motivated by two experimental observations during snowfalls: crystal aggregation at the snow surface \cite{grl}, and spatial distribution of snowflakes in air \cite{NC}. To interpret these observations, we propose a simple but general theory explaining how particles are able to keep aggregating and growing in turbulent flows, while modifying their trajectory and their response to local velocity fluctuations. The fundamental requirements for particle aggregation are i) the occurrence of collisions and ii) the ability of approaching particles to stick together, bond, or coalesce. Among many examples of particle growth by aggregation mechanisms, we describe in details the case of snowflake formation and then extend the discussion to droplets and other inertial particles.

The aggregation of snow crystals is fairly well understood at the snow surface. On flat terrain, during snowfall under negligible wind conditions, settling crystals and snowflakes are observed to form surface roughness structures much larger than the crystal diameter \citep[][]{grl}. This phenomenon was revealed by observing that the temporal increase of snow  roughness during a snowfall follows a power law  with the characteristic KPZ exponent of interface growth \citep[][]{KPZ}. This result has been further confirmed through simulations of ballistic deposition with particles obeying to a strict sticking rule \cite{manes}, which suggests that snow crystals can form large surface aggregates only if they bond to each other as soon as they hit the snow surface. Micro-meteorological conditions, such as relatively warm temperature and high relative humidity, play a major role in the bonding of ice crystal, allowing mechanical interlocking or sintering,  without fracture of either the impacting crystals, or those forming the upper layer of the snow cover.

In air, the growth of snowflakes may follow a similar path. If the right humidity and temperature conditions are met, colliding snow crystals may form a snowflake aggregate.  However, while collisions on the snow cover are a direct consequence of snow settling, collisions in air depend on both the inertial properties of the snowflakes, and on the characteristics of the turbulent flow. Thus, both micro-meteorology (enabling aggregation) and particle-turbulence interactions (enabling collisions) are important in the growth of snowflakes in air. Qualitatively, this is supported by the field observation that the size of snowflakes, and thus their inertial properties, can drastically change within the same snowfall event, even under similar temperature and relative humidity conditions (e.g. \cite{theriault}). This implies that the physical processes governing  snowflake size, i.e. the aggregation mechanism in air, can evolve within relatively short time scales, of the order of the integral scale of atmospheric turbulence, in addition to relatively long meteorological time scales.
The significant variability of atmospheric turbulent conditions is reflected in the wide range of observed snowflake sizes, from O$(10^{-4})$ m to O$(10^{-2})$ m, resulting from the aggregation of a different number of dendritic crystals \cite{lang,Libbrecht2005,garr}.

In this work we propose a mechanistic model, based on particle-turbulence interaction mechanisms and inertial particle clustering \cite{bala}, able to predict the maximum size of snow crystal aggregates and explain the variability in snowflake size observed in nature. The theory is generalized to any type of inertial particles, immersed in a homogeneous turbulent flow, and able to bond during a collision as a result of, e.g. gravitational, electrostatic, or chemical  forces \citep{guasto,cuzzi}.

\textbf{Inertial particle clustering in turbulent flows.} As a first working hypothesis, we assume here that the key mechanism controlling particle aggregation, once particles have the ability to stick together, is the same one responsible for inertial particle preferential concentration \cite{bala}. The optimal conditions for clustering of inertial particles, neglecting the effect of gravity, are realized when the Stokes number is critical, i.e. when $St=\tau_p/\tau_f=1$, where $\tau_p$ is the particle response time and $\tau_f$ is a representative flow time scale (e.g. \cite{eaton}). This definition of the Stokes number is derived under the assumption that drag and inertia are the only terms in the particle equation of motion \citep{maxey}.

There is no ambiguity in the definition of the particle response time, even for particles somewhat larger than the Kolmogorov scale, as long as the drag remains the dominant force (e.g. \cite{burton}). Assuming Stokes drag, with the options to correct for particle Reynolds number \cite{clark}, the particle response time is defined as $\tau_p=\frac{\rho_s/\rho D^2}{18\nu}$, where $\rho_s$ is the particle density (much larger than the fluid density $\rho$), $D$ is the equivalent particle diameter, and $\nu$ is the fluid kinematic viscosity. Note that when $\rho_s\simeq \rho$, other terms from the particle equation of motion, e.g. the added mass or the Basset force, must be accounted for \cite{bala}.

The flow time scale $\tau_f$ is not trivial to define. For particles smaller than the Kolmogorov length scale ($\eta$), numerical and experimental results suggest $\tau_f=\tau_\eta$, the Kolmogorov time scale, based on the observation that clustering indeed occurs when $St =\tau_p/\tau_\eta \simeq 1$ \cite{ferrante,hogan}. For particles larger than the Kolmogorov scale, the question is more complex (e.g. \cite{variano,fiabane}). On the one hand, even the smallest eddies can modify the flow field at the interface between the fluid and the particle, and therefore contribute, though minimally, to the pressure field and the resulting lift and drag controlling the particle motion. On the other hand, such large particle will not orbit, or be preferentially swept around Kolmogorov eddies. This implies that the dominant flow time scale that such large particle will experience must be larger than $\tau_\eta$, and smaller than the integral scale of the flow.

\begin{figure}
\includegraphics[width=7cm]{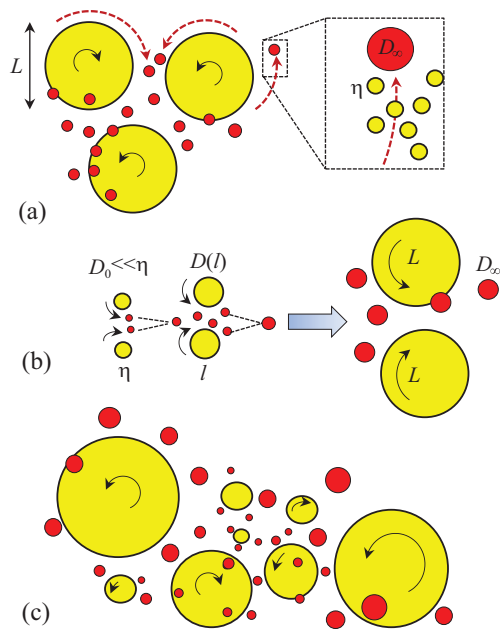}
  \caption{Schematics of dynamic particle-turbulence interactions: a) Large (red) particle aggregates do not respond to Kolmogorov scales, and rather cluster in strain dominated regions between large (yellow) eddies; b) Particle growth from sub-Kolmogrov scale to maximum size $D_\infty$ under dynamically critical conditions and energy containing eddies of integral scale $L$; c) Distribution of particle size around turbulent eddies of varying length scale.}\label{ske}
\end{figure}

We now describe how we estimate $\tau_f$ for a particle of increasing size, such as a snowflake or a droplet.
Guala et al. \cite{guala} observed that relatively large silica gel particles in water with $D \simeq 5\eta$ and $\rho_s/\rho\simeq 1.4$, were preferentially concentrated in high strain regions, in between eddies whose mean radius of curvature ($r$) was on the order of the Taylor microscale $\lambda \simeq 10\eta$. Particles were also observed to follow closer orbits, when conditioned on locally high vorticity, as well as cluster in high strain regions in between such Taylor-size vortices. This observation raised the question on if and how a group of particles, with a wide range in size, would distribute around corresponding turbulent eddies of varying scales.  Particles of different sizes would perceive different flow time scales, depending on the spatial variability of the velocity field within a domain defined by the particle trajectory and, thus, by its curvature radius $r$, (see Fig. \ref{ske}a).
Under stationary, homogeneous, isotropic turbulence, for a particle moving along a perfectly round trajectory of length $l=\pi r$, we can define the flow time scale perceived by the particle as a $\tau_f(l)=l/\delta(u(l))$ \cite{frish}, where $\delta u(l)=\langle u(x+l)-u(x) \rangle$ is the scale-dependent velocity increment  (averaged along a generical $x$ direction of the velocity component).
Such flow time scale $\tau_f(l)$ represents a scale-dependent eddy turnover time. It can be extended to more complex orbits and implemented in the definition of Stokes number to assess which particles would become critical ($ \tau_p/\tau_f \simeq 1$).

\textbf{Governing equations and predictive analysis for particle growth.} Our  main working hypothesis is that, while particles collide, stick together, and grow in size, their Stokes number remains critical.  Phenomenologically this implies that progressively larger particles will cluster, collide and aggregate around progressively larger eddies. We start with the introduction of an initial state of small particles ($D_0<\eta$) in critical conditions, i.e. $St(D_0)=1$. Obviously, if the initial state does not favor clustering,  particle collisions would be rare and particle size growth would not occur, in which case our model would not apply. However, if clustering occurs, the evolution of particle size under such dynamically-critical Stokes conditions can be described as follows.

The initial state particle response time is estimated as $\tau_{p0}=\frac{D_0^2\rho_p/\rho}{18\nu} \propto D^2_0$. The flow time scale perceived by the particle can be expressed  as a function of the eddy size around which it orbits: $\tau_f(l)\sim l/\delta u \sim \epsilon^{-1/3}l^{2/3}$, where $\epsilon$ is the turbulent dissipation rate and $\tau_f$ is assumed to vary within the turbulent inertial range, i.e. where  $\delta u \propto \epsilon^{1/3}l^{1/3}$ applies \cite{frish}.
Imposing $St=\frac{\tau_p(D)}{\tau_f(l)}=1$, we obtain an equation for the evolution of the particle diameter under dynamically critical conditions:
\begin{equation}\label{eq:1}
D^2(l)=\frac{18 \nu \epsilon^{-1/3}l^{2/3}}{\rho_p/\rho}.
\end{equation}
Keeping $St=1$ implies that, as the particle size increases, also the relevant flow timescale $\tau_f$ increases, within the inertial range, from the Kolmogorov time scale $\tau_\eta$, to the integral time scale $T_i=L/u_{rms}$ ($L$ is the integral length scale and $u_{rms}$ is the r.m.s. velocity), as sketched in Fig. \ref{ske}b.  The lower and upper ends of the turbulence inertial range, applied to equation \ref{eq:1}, define two limiting cases. The first is at the earliest state of particle growth $D_0$. For $l\simeq \eta$, and critical Stokes number $St=1$, we must recover the initial conditions favoring clustering. From $D^2_0=\frac{18 \nu \epsilon^{-1/3}\eta^{2/3}}{\rho_p/\rho}$, we replace $\eta= (\nu^3/\epsilon)^{1/4}$, leading to
$D^2_0=\frac{18 \nu \tau_\eta}{\rho_p/\rho}$ with  $\tau_\eta=(\nu/\epsilon)^{1/2}=\tau_p$.
This confirms that small particles, i.e. $D_0<<\eta$, behave as predicted by the typical Kolmogorov-based Stokes number $St_{\eta}=\tau_p/\tau_{\eta}$. %Because critical conditions are established at the initial state, clustering is expected to occur, while particles start colliding, growing in size and orbiting around turbulent eddies.

The second limiting case is the maximum particle size ($D_{\infty}$) reached when $l \simeq L$. The integral length scale $L$ represents the largest orbit around which particles can be trapped, or preferentially swept, at their latest evolutionary state.
From $D^2_{\infty}=\frac{18 \nu \epsilon^{-1/3}L^{2/3}}{\rho_p/\rho}$,
we can express the dissipation rate $\epsilon=u_{rms}^3/L$ leading to
\begin{equation}\label{eq:3}
    D^2_{\infty}=\frac{18 \nu }{\rho_p/\rho}L/u_{rms}
\end{equation}
Equation (\ref{eq:3}) allows to predict the maximum aggregate size, under dynamically critical Stokes conditions (see Fig. \ref{ske}b). At any scale larger than $L$, the velocity would be de-correlated and no eddies $l>L$ would exist to trap the particles. This implies that the turbulent structure with time scale $\tau_f$ necessary to keep $St=1$ would not be present in the flow, hence particles of size $D_\infty$ stop clustering, colliding and growing in size. Therefore, $D_\infty$ is the largest particle size resulting from size-dependent clustering, collisions and aggregation, in a cascade of critical Stokes conditions occurring from $\eta$ to $L$. %$\tau_f=\tau_{\eta}$ to $\tau_f=L/u$.
Representing dissipation in terms of the energy-containing-eddies provides an alternative way to estimate $D_\infty$ through measurable quantities such as $L$ and $u_{rms}$, characterizing the stationary isotropic turbulence. To highlight the dependency between the maximum particle size and the Reynolds number of the ambient flow ($Re=u_{rms} L/\nu$), we express the ratio of Kolmogorov scale to integral scale as $\eta/L=Re^{-3/4}$. This leads to the alternative expression for $D_\infty$:
\begin{eqnarray}\label{eq:4}
      \left( \frac{D_\infty}{\eta}\right)^2=\frac{18 Re^{1/2}}{\rho_p/\rho}.
\end{eqnarray}
Equation (\ref{eq:4}) illustrates how a turbulent flow at higher $Re$, with a wider inertial range, can sustain a larger range of particle sizes (the upper limit $D_\infty$ grows).

By imposing a dynamically critical Stokes number we are considering the most favorable conditions for particle aggregation and size growth. However, not all particles in the flow will necessarily reach the limiting particle size. At any time, we expect to observe a distribution of particle sizes, resulting from the distribution of eddies of different size characteristic of turbulent flows: small particles orbiting around Kolmogorov scale eddies, coexisting with larger particles clustering in between larger and more energetic eddies, as skecthed in Fig. \ref{ske}c. To account for the occurrence of Stokes critical conditions for a generic particle of size $D$, orbiting around eddy of scale $l$ and flow time scale $\tau_f(l)=\tau_p(D)$, we rewrite equation (\ref{eq:1}), maintaining explicit the dependency on $l$:
\begin{eqnarray}\label{eq:5}
      \frac{D(l)}{\eta}=\left( \frac{18 \nu^{-1/2} \epsilon^{1/6}}{\rho_p/\rho}\right)^{1/2}  l^{1/3}.
\end{eqnarray}

Equation \ref{eq:5} predicts that the particle size grows as $l^{1/3}$, in the inertial range scaling regime  ($\eta<l<L$) where the scale-dependent turbulent kinetic energy is known to increases as $l^{5/3}$. The predicted power-law particle growth regime cannot be extended beyond the turbulent inertial range (i.e. for scales larger than $L$), because $D \rightarrow D_\infty$ as $l  \rightarrow L$.

Note that, in the above framework, particles do not exert any feedback on the flow (one-way coupling), and there is no direct dependency on time. The phenomenology of particle growth by progressive self-organization of particle clusters, would require a Lagrangian time scale \cite{toschi}. However, if we interpret $D$ as an average particle diameter, then an Eulerian time scale would be more appropriate to describe the statistical evolution of the particle ensemble, as a function of, e.g. collision or coalescence probability, and initial particle concentration. In the simplest approach followed here we bypass the temporal evolution, which requires an environment-specific or particle-specific aggregation law, and describe the particle size in terms of initial ($D_0$), intermediate ($D(l)$) and saturation ($D_\infty$) states.

\begin{figure}
\includegraphics[width=8.5cm]{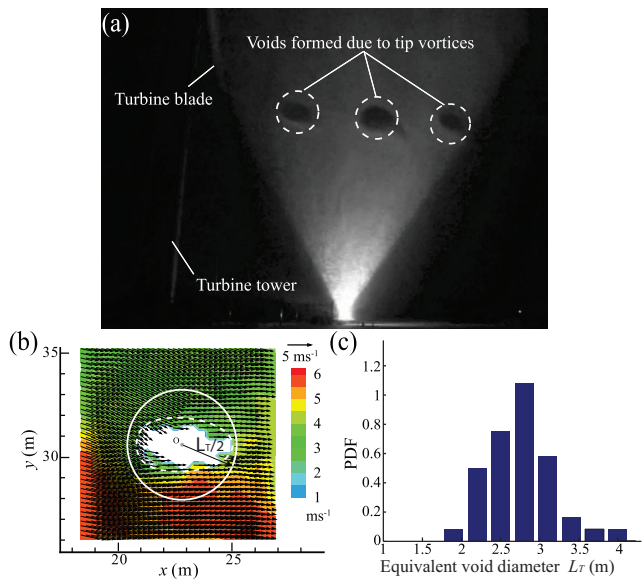}
  \caption{Field measurements: a) Picture of the tip vortex voids in the wind turbine wake; b) Velocity distribution around a sample of snow void; c) size distribution of snow void diameters $L_T$ observed during the field deployment, leading to $\langle L_T \rangle =2.8$ m.}\label{fig:vortex1}
\end{figure}

\textbf{Evidence of large-scale particle preferential-concentration in a wind turbine wake.} Fig. \ref{fig:vortex1}a shows the instantaneous distribution of snowflakes around the tip vortex generated in the wake of a 2.5 MW turbine, as resolved by our Super-Large-Scale Particle Image Velocimetry measurements \cite{NC}. The observed macroscopic regions of low snowflake concentration (voids) are the result of snowflakes-turbulence interaction in the core of the tip vortex. The latter, generated at the tip of each wind turbine blade, has a well defined length scale and spatial structure: it is a coherent vortex tube with a core diameter $L_T$, scaling  with the blade chord length, and much larger than any snowflake. The tip vortex  is characterized by a vorticity field strong enough to induce and maintain, in a Lagrangian sense, a strong particle preferential concentration, as manifested in the traveling voids of Fig \ref{fig:vortex1}a. Flow measurements in the area surrounding the tip vortex (Fig. \ref{fig:vortex1}b) allowed the estimate of the circulation  $\Gamma=\oint \overrightarrow{u}dl$, which we use to derive the tip vortex flow time scale $\tau_f=\pi L_{T}^2/\Gamma$. In these specific turbine operating conditions, we identified a number of voids in the illuminated turbine wake,  estimated their size $L_T$, and plotted their distribution (see Fig. \ref{fig:vortex1}c). With averaged $L_T=2.8$ m and $\Gamma= 7.6$ m$^2$s$^{-1}$ \cite{NC}, the mean tip vortex timescale is $\tau_f=3.2$ s.
The average snowflake density and size are estimated as $\rho_p \approx 24.5$ Kg m$^{-3}$ and $D=2\cdot 10^{-3}$, respectively, from measurements of vertically-averaged settling velocities \citep[see][for details]{NC}.
For such density, about 50\% lighter than fresh snow on the ground \cite{andy}, the critical conditions for the tip vortex flow ($\tau_p/\tau_f=1$) occur for snowflakes of size $D_{LT}=6.6\cdot 10^{-3}$ m, about three times larger than the estimated average diameter. The resulting range of snowflake size might well be the cause of the observed vertical variability in snow settling velocity, from  0.5 to 1.2 ms$^{-1}$ \cite{NC}, which could not be explained by the dynamics of the turbine wake. While rigorous proof is still needed, Fig. \ref{fig:vortex1} provides experimental evidence that particle larger than the Kolmogorov scale can be preferentially swept by large scale vortices, at the edges of which particle collisions and size growth may occur. We do not have experimental validation on the snowflake aggregate size at the edge of the tip vortex, but we clearly observe that no snowflakes accumulate in the vortex core. We suggest that such mechanism of large scale preferential sweeping  may also occur under specific strong turbulent conditions in clouds. The latter provide a more homogeneous turbulent environment in which our predictions can be further tested.
%Fig \ref{fig:vortex1}

\textbf{Droplet and snowflake formation.}
It is conceivable that in a cloud, characterized by a wide range of turbulent scales, a subset of droplets with a specific size will start to cluster. Clustering enhances the probability of mutual collisions and coalescence, favoring droplet growth and eventually precipitation. The large Reynolds number, and scale separation between $\eta$ and $L$, and thus between $D_0$ and $D_{\infty}$.  Considering a typical stratocumulus cloud with turbulent length and velocity scales $L\simeq 100$ m and $u_{rms}\simeq 1$ ms$^{-1}$, and $\rho_p/\rho=10^3$ (estimates consistent with direct measurements \cite{shaw10,lothon}), we obtain $D_\infty \simeq 5$ mm, which agrees well with the large droplet size observed in thunderstorms. For the lower size case limit, setting $\tau_\eta=0.025$ s following \cite{shaw10}, we obtain $D_0=50$ $\mu$m, which is reasonably close to the initial, critical droplet size \cite{shaw12}. To clarify, $\tau_\eta$ was estimated using the second order structure function from high frequency velocity measurements in clouds, leading to $\epsilon=2.9\cdot 10^{-4}$ m$^2$s$^{-3}$ \cite{shaw10}.
In the case of ice crystals, aggregating into snowflakes under the same cloud turbulence conditions ($L\simeq 100$ m and $u_{rms}\simeq 1$ ms$^{-1}$), the average snowflake density in fact determines $D_\infty$. With $\rho_p / \rho \approx 20$ (weakly depending on air temperature and relative humidity), we obtain $D_\infty\approx 36$ mm, which matches the upper limit of the range of typically observed snowflake sizes \cite{lang,garr}.

We acknowledge, as significant model limitations, that: i) the particle density remains constant during particle growth, ii) collisions are only promoting particle aggregation, without contributing to size depletion, as e.g. snowflake structural collapse or break-up; iii) size growth due to differential settling is not accounted for, implying that snowflakes or droplet size observed at ground may be under-predicted.

We conclude that the mechanism of size growth due to particle aggregation in dynamically-critical Stokes conditions is able to i) explain the variability in droplet size, snowflake size and hydrometeor settling velocity observed in nature \cite{garr,shaw10,NC}, and ii) predict the maximum size of particle aggregates in turbulent flows.

\end{document}